\documentclass[11pt, svgnames, twocolumn]{article}
\usepackage[switch, columnwise]{lineno} 
\usepackage{amsmath, amsthm, amssymb, amsfonts, bbm, graphicx}
\usepackage[top=1in, left=0.75in, bottom=1in, right=0.75in]{geometry}
\usepackage[lastpage,user]{zref}
\usepackage{fancyhdr}
\usepackage[sort, numbers]{natbib}
\usepackage{mdframed}
\usepackage{subcaption}

\usepackage{dsfont}

\usepackage{color, xcolor, fancybox, ascii, booktabs, colortbl, enumerate}

\graphicspath{{./figures/}{./figures_old/}}
\newcolumntype{a}{>{\columncolor{black!10}}c}

\newcommand{\E}{\mathbb{E}}

\renewcommand{\P}{\mathbb{P}}

\newcommand{\ppv}{\mathrm{PPV}}
\newcommand{\fpr}{\mathrm{FPR}}
\newcommand{\fnr}{\mathrm{FNR}}

\numberwithin{equation}{section}

\theoremstyle{plain}
    
    \newtheorem*{theorem*}{Theorem}
    \newtheorem{prop}{Proposition}
    \newtheorem{corollary}{Corollary}
\theoremstyle{definition}
    \newtheorem{definition}{Definition}
    \newtheorem*{definition*}{Definition}
    
\numberwithin{theorem}{section}
\numberwithin{prop}{section}
\numberwithin{corollary}{section}
\numberwithin{definition}{section}

\title{\vspace{-1cm} Fair prediction with disparate impact: \\ {\Large A study of bias in recidivism prediction instruments}}
\author{Alexandra Chouldechova \vspace{0.5em} \\
Heinz College, Carnegie Mellon University\\
5000 Forbes Avenue, Pittsburgh, PA, USA  \\
{\tt achould@cmu.edu}
}

\date{}

\begin{document}
  \maketitle
  
  
  \begin{abstract}
 Recidivism prediction instruments (RPI's) provide decision makers with an assessment of the likelihood that a criminal defendant will reoffend at a future point in time. While such instruments are gaining increasing popularity across the country, their use is attracting tremendous controversy. Much of the controversy concerns potential discriminatory bias in the risk assessments that are produced. This paper discusses a fairness criterion originating in the field of educational and psychological testing  that has recently been applied to assess the fairness of recidivism prediction instruments. We demonstrate how adherence to the criterion may lead to considerable disparate impact when recidivism prevalence differs across groups. 
  \end{abstract}
  
  \vspace{-0.25em}
  
\section{Introduction} \label{sec:intro}

\vspace{-0.25em}

Risk assessment instruments are gaining increasing popularity within the criminal justice system, with versions of such instruments being used or considered for use in pre-trial decision-making, parole decisions, and in some states even sentencing \citep{blomberg2010validation, 538rpi}.  In each of these cases, a high-risk classification---particularly a high-risk misclassification---may have a direct adverse impact on a criminal defendant's outcome.    If RPI's are to continue to be used, it is important to ensure that they do not result in unethical practices that disparately affect different groups.

Within the psychometrics literature, there exist widely accepted and adopted standards for assessing whether an instrument is fair in the sense of being free of \emph{predictive bias}.  These standards have recently been applied to the COMPAS \cite{compasfaq} and PCRA \cite{pcra} instruments, with initial findings suggesting that there is evidence of predictive bias when it comes to gender, but not when it comes to race \cite{singh2013predictive, skeem2015risk, skeem2016gender}.  

In a recent widely popularized investigation of the COMPAS RPI conducted by a team at ProPublica, a different approach to assessing instrument bias told what appears to be a contradictory story \cite{propublica2016}.  The authors found that the likelihood of a non-recidivating Black defendant being assessed as high-risk is nearly twice that of White defendants.  While this analysis has met with much criticism, it has also made headlines. There is no doubt that it is now embedded in the national conversation on the use of RPI's.

In this paper we show that the differences in false positive and false negative rates cited as evidence of racial bias in the ProPublica article are a direct consequence of applying an instrument that is free from predictive bias\footnote{in the psychometric sense} to a population in which recidivism prevalence differs across groups.   Our main contribution is twofold.  (1) First, we make precise the connection between the psychometric notion of test fairness and error rates in classification.  (2) Next, we demonstrate how using an RPI that has different false postive and false negative rates between groups can lead to disparate impact when individuals assessed as high risk receive stricter penalties.  Throughout our discussion we use the term \emph{disparate impact} to refer to settings where a penalty policy has unintended disproportionate adverse impact on a particular group.

It is important to bear in mind that fairness itself---along with the notion of disparate impact---is a social and ethical concept, not a statistical one.  An instrument that is free from predictive bias may nevertheless result in disparate impact depending on how and where it is used.  In this paper we consider hypothetical use cases in which we are able to directly connect statistically quantifiable features of RPI's to a measure of disparate impact.

\vspace{-0.25em}
\subsection{Data description and setup} \label{sec:intro_outline}

The empirical results in this paper are based on the Broward County data made publicly available by ProPublica \cite{propublica2016data}.  This data set contains COMPAS recidivism risk decile scores, 2-year recidivism outcomes, and a number of demographic and crime-related variables.    We restrict our attention to the subset of defendants whose race is recorded as African-American ($b$) or Caucasian ($w$). 
  
  \vspace{-0.25em}
\section{Assessing fairness}
\vspace{-0.25em}

We begin by with some notation.  Let $S = S(x)$ denote the risk score based on covariates $X = x$, with higher values of $S$ corresponding to higher levels of assessed risk.  Let $R \in \{b, w\}$ denote the group that the individual belongs to, which may be one of the components of $X$.  Lastly, let $Y \in \{0 , 1\}$ be the outcome indicator, with $1$ denoting that the given individual recidivates.  
  In this notation, we can think of the psychometric test fairness condition roughly as follows. 
 
 \begin{definition}[Test fairness]
   A score $S = S(x)$ is  \emph{test-fair} (well-calibrated)\footnote{Depending on the context, we may further desire that this criterion is satisfied when we condition on some of the covariates.  Our analysis extends to this case as well.}   if it reflects the same likelihood of recidivism irrespective of the individual's group membership, $R$. That is, if for all values of $s$,
   \begin{equation}
   \P(Y = 1 \mid S = s, R = b) = \P(Y = 1 \mid S = s, R = w).
   \label{def:testfair}
   \end{equation}
   
 \end{definition}
 
\begin{figure}[t]
    \centering
    \includegraphics[width = \linewidth]{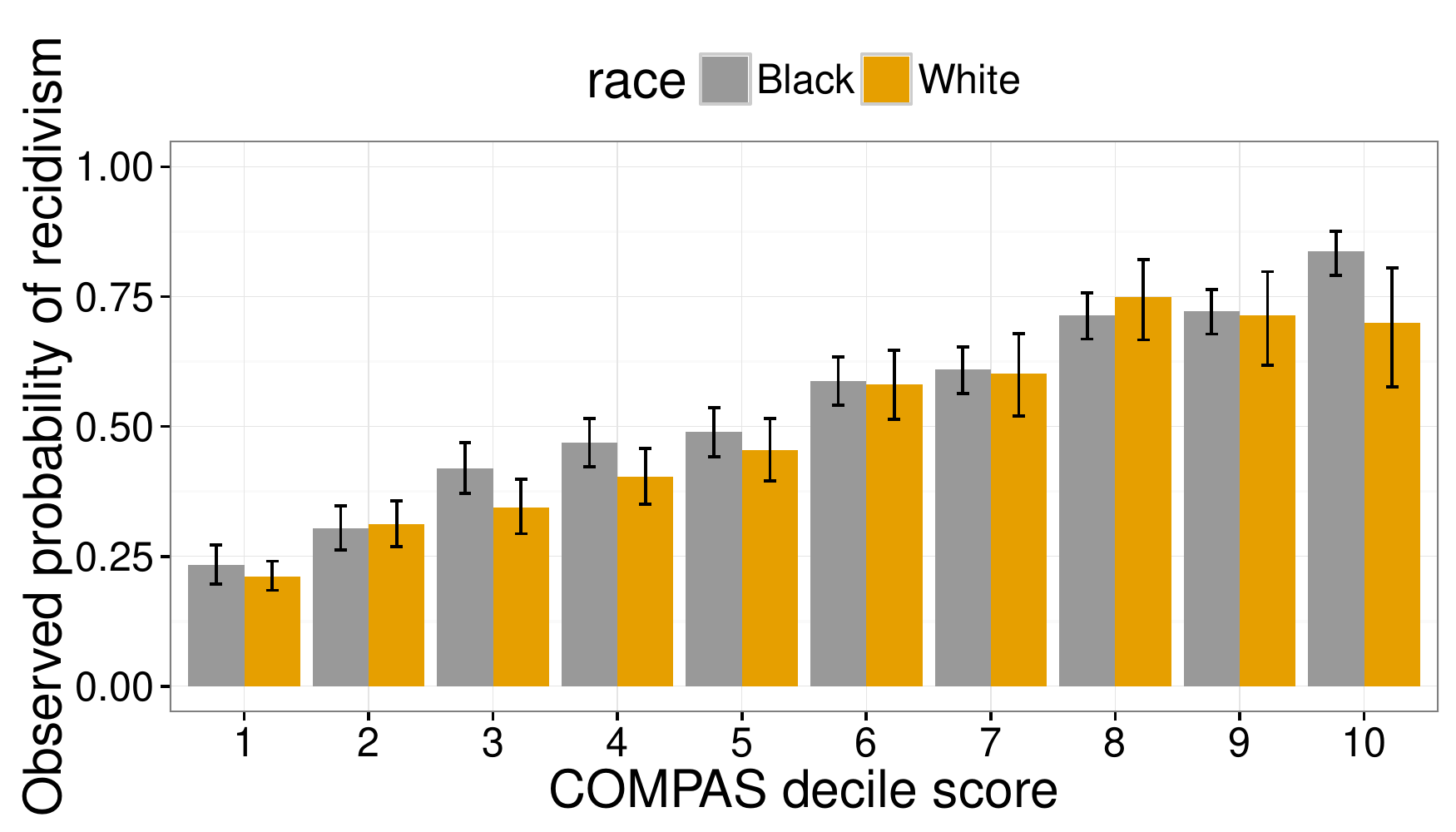}
    \caption{\small{Plot shows $\P(Y = 1 \mid S = s, R)$ for the COMPAS decile score, with $R \in \{\text{Black}, \text{White}\}$.  Error bars represent 95\% confidence intervals.}}
    \label{fig:ppv_plot}
 \end{figure}

 Figure~\ref{fig:ppv_plot} shows a plot of the observed recidivism rates across all possible values of the COMPAS score.  We can see that the COMPAS RPI appears to adhere well to the test fairness condition.   In their response to the ProPublica investigation, \citet{floresfalse} further verify this adherence using logistic regression.
 
 \vspace{-0.5em}
 \subsection{Implied constraints on the false positive and false negative rates}
 
 To facilitate a simpler discussion of error rates, we introduce the \emph{coarsened score} $S_c$, which is obtained by thresholding $S$ at some cutoff $s_{HR}$. 
 \begin{equation}
   S_c(x) \equiv \begin{cases}
   \text{HR} & \text{if } S(x) > s_{HR} \\
   \text{LR} & \text{if } S(x) \le s_{HR}
   \end{cases}
 \end{equation}
 
 The coarsened score simply assesses each defendant as being at \emph{high-risk} or \emph{low-risk} of recidivism.  For the purpose of our discussion, we will think of $S_c$ as a classifier used to predict the binary outcome $Y$.  This allows us to summarize $S_c$ in terms of a confusion matrix, as shown below.
 
     \begin{tabular}{|r|rr|}
       \hline
      & $S_c = $ Low-Risk & $S_c = $ High-Risk \\ 
       \hline
     $Y = 0$ & TN & FP \\ 
       $Y = 1$ & FN & TP \\ 
        \hline
     \end{tabular} \\

 It is easily verified that test fairness of $S$ implies that the \emph{positive predictive value} of the coarsened score $S_c$ does not depend on $R$.  More precisely, it implies that that the quantity
\begin{equation}
  \ppv(S_c \mid R = r) \equiv \P(Y = 1 \mid S_c = \text{HR}, R = r)
  \label{eq:equiv_ppv}   
\end{equation}
does not depend on $r$.
Equation \eqref{eq:equiv_ppv} thus forms a necessary condition for the test fairness of $S$.  We can think of this as a constraint on the values of the confusion matrix.  A second constraint---one that we have no direct control over---is the recidivism prevalence within groups, which we denote here by $p_r \equiv \P(Y = 1 \mid R = r)$.  

Given values of the $\ppv \in (0,1)$ and prevalence $p \in (0,1)$, it is straightforward to show that the \emph{false negative rate} \mbox{$\fnr = \P(S_c = \text{LR} \mid Y = 1)$} and \emph{false positive rate} \mbox{$\fpr = \P(S_c = \text{HR} \mid Y = 0)$} are related via the equation
\begin{equation}
  \fpr = \frac{p}{1 - p} \frac{1 - \ppv}{\ppv} ( 1 - \fnr).  
  \label{eq:fpr_fnr}
\end{equation}
A direct implication of this simple expression is that when the recidivism prevalence differs between two groups, a test-fair score $S_c$ cannot have equal false positive and negative rates across those groups.\footnote{This observation is also made in independent concurrent work by \citet{kleinberg2016inherent}. }  
  
This observation enables us to better understand why the ProPublica authors observed  large discrepancies in FPR and FNR between Black and White defendants.\footnote{Black: FPR = 45\%, FNR = 28\%. \\  White: FPR = $23\%$, FNR = $48\%$ }  The recidivism rate among back defendants in the data is 51\%, compared to 39\% for White defendants.  Since the COMPAS RPI approximately satisfies test fairness, we know that some level of imbalance in the error rates must exist.

\section{Assessing impact}

\vspace{-0.25em}

In this section we show how differences in false positive and false negative rates can result in disparate impact under policies where a high-risk assessment results in a stricter penalty for the defendant.  Such situations may arise when risk assessments are used to inform bail, parole, or sentencing decisions.  In the state of Pennsylvania, for instance, statutes permit the use of RPI's in sentencing, provided that the sentence ultimately falls within accepted guidelines.  We use the term ``penalty'' somewhat loosely in this discussion to refer to outcomes both in the pre-trial and post-conviction phase of legal proceedings.  Even though pre-trial outcomes such as the amount at which bail is set are not \emph{punitive} in a legal sense, we nevertheless refer to bail amount as a ``penalty'' for the purpose of our discussion. 

There are notable cases where RPI's are used for the express purpose of informing risk reduction efforts.  In such settings, individuals assessed as high risk receive what may be viewed as a benefit rather than a penalty.  The PCRA score, for instance, is intended to support precisely this type of decision-making at the federal courts level.  Our analysis in this section specifically addresses use cases where high-risk individuals receive stricter penalties.

To begin, consider a setting in which guidelines indicate that a defendant is to receive a penalty $t_L \le T \le t_H$.  A very simple risk-based approach, which we will refer to as the MinMax policy, would be to assign penalties as follows:
\begin{equation}
 T_{\mathrm{MinMax}} =  \begin{cases}
      t_L & \text{if } S_c =  \text{ Low-risk} \\
      t_H & \text{if } S_c =  \text{ High-risk}
      \end{cases}.
\end{equation}

In this simple setting, we can precisely characterize the extent of disparate impact in terms of recognizable quantities. Define $T_{r,y}$ to be the penalty given to a defendant in group $R = r$ with observed outcome $Y=y \in\{0,1\}$, and let $\Delta = \Delta(y_1, y_2) = \E(T_{b,y_1} - T_{w,y_2})$ be expected difference in sentence between defendants in different groups.  $\Delta$ is a measure of disparate impact.

\begin{prop}
  The expected difference in penalty under the MinMax policy is given by
  \begin{align*}
  \Delta &\equiv \E_\mathrm{MinMax} (T_{b,y_1}  - T_{w,y_2}) \\ 
  &= (t_H- t_L)\big[\P(S_c = \text{HR} \mid R = b, Y = y_1)  \\
  & \hspace{7em} - \P(S_c = \text{HR} \mid R = w, Y = y_2)\big]
  \end{align*}
\end{prop}
\noindent We will discuss two immediate Corollaries of this result.
\begin{corollary}[Non-recidivators]
  Among individuals who \emph{do not recidivate}, the difference in average penalty under the MinMax policy is 
  \begin{equation}
    \Delta = (t_H - t_L)(\fpr_b - \fpr_w)
    \label{eq:survive}
  \end{equation}
\end{corollary}
\begin{corollary}[Recidivators]
  Among individuals who \emph{recidivate}, the difference in average penalty under the MinMax policy is 
  \begin{equation}
    \Delta = (t_H - t_L)(\fnr_w - \fnr_b)
    \label{eq:recid}
  \end{equation}
\end{corollary}

When using a test-fair RPI in populations where recidivism prevalence differs across groups, it will generally be the case that the higher recidivism prevalence group will have a higher FPR and lower FNR.  From equations~\eqref{eq:survive} and \eqref{eq:recid}, we can see that this would result in greater penalties for defendants in the higher prevalence group, both among recidivating and non-recidivating offenders.   

An interesting special case to consider is one where $t_L = 0$.  This could arise in sentencing decisions for offenders convicted of low-severity crimes who have good prior records.  In such cases, so-called restorative sanctions may be imposed as an alternative to a period of incarceration. If we further take $t_H = 1$, then $\E T = \P(T \neq 0)$, which can be interpreted as the probability that a defendant receives a sentence imposing some period of incarceration.  

It's easy to see that in such settings a non-recidivating defendant in group $b$ is $\fpr_b /  \fpr_w$ times more likely to be incarcerated compared to a non-recidivating defendant in group $w$.\footnote{We are overloading notation in this expression:  Here, \mbox{$\fpr_r = \P(\mathrm{HR} \mid R = r, t_L = 0)$}, similarly for $\fnr_r$.}  This naturally raises the question of whether overall differences in error rates are observed to persist across more granular subgroups.

\begin{figure}[t]
  \centering
  \includegraphics[width = \linewidth]{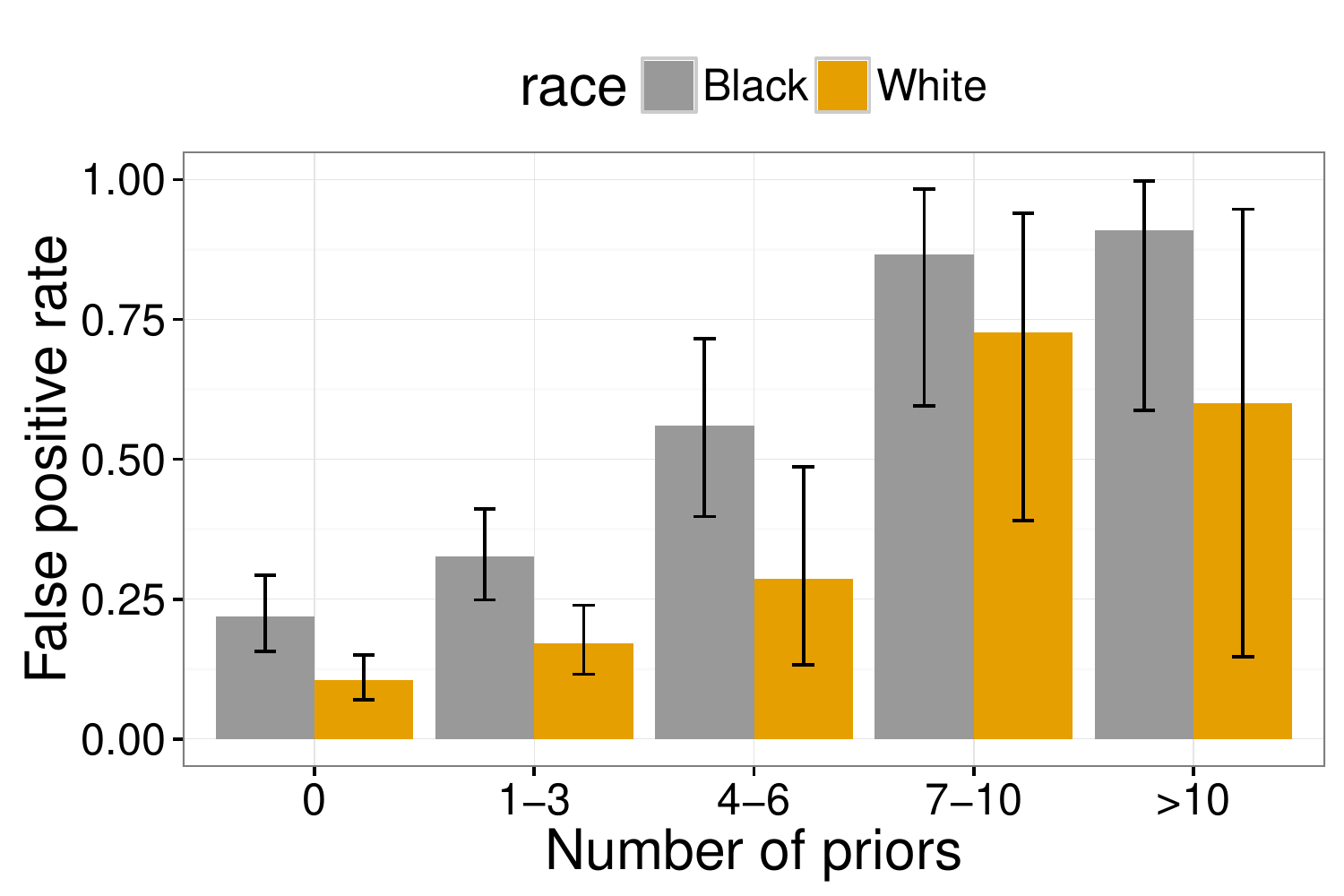}
  \caption{\small False positive rates across prior record count for defendants charged with a Misdemeanor offense.  Plot is based on assessing a defendant as ``high-risk'' if their COMPAS decile score is $>4$. Error bars represent 95\% confidence intervals.}
  \label{fig:fpr_prior}
\end{figure}

One might expect that differences in false positive rates are largely attributable to the subset of defendants who are charged with more serious offenses and who have a larger number of prior arrests/convictions.  While it is true that the false positive rates within both racial groups are higher for defendants with worse criminal histories, considerable between-group differences in these error rates persist across low prior count subgroups.  Figure~\ref{fig:fpr_prior} shows a plot of false positive rates across different ranges of prior count for defendants charged with a misdemeanor offense, which is the lowest severity criminal offense category.  As one can see, differences in false positive rates between Black defendants and White defendants persist across prior record subgroups.

\vspace{-0.5em}
\subsection{Connections to measures of effect size}

  A natural question to ask is whether the level of disparate impact, $\Delta$, is related to some measures of effect size commonly used in scientific reporting.  With a small generalization of the \emph{\% non-overlap} measure, we can answer this question in the affirmative.  
  
\begin{figure}[t]
    \centering
    \includegraphics[width = \linewidth]{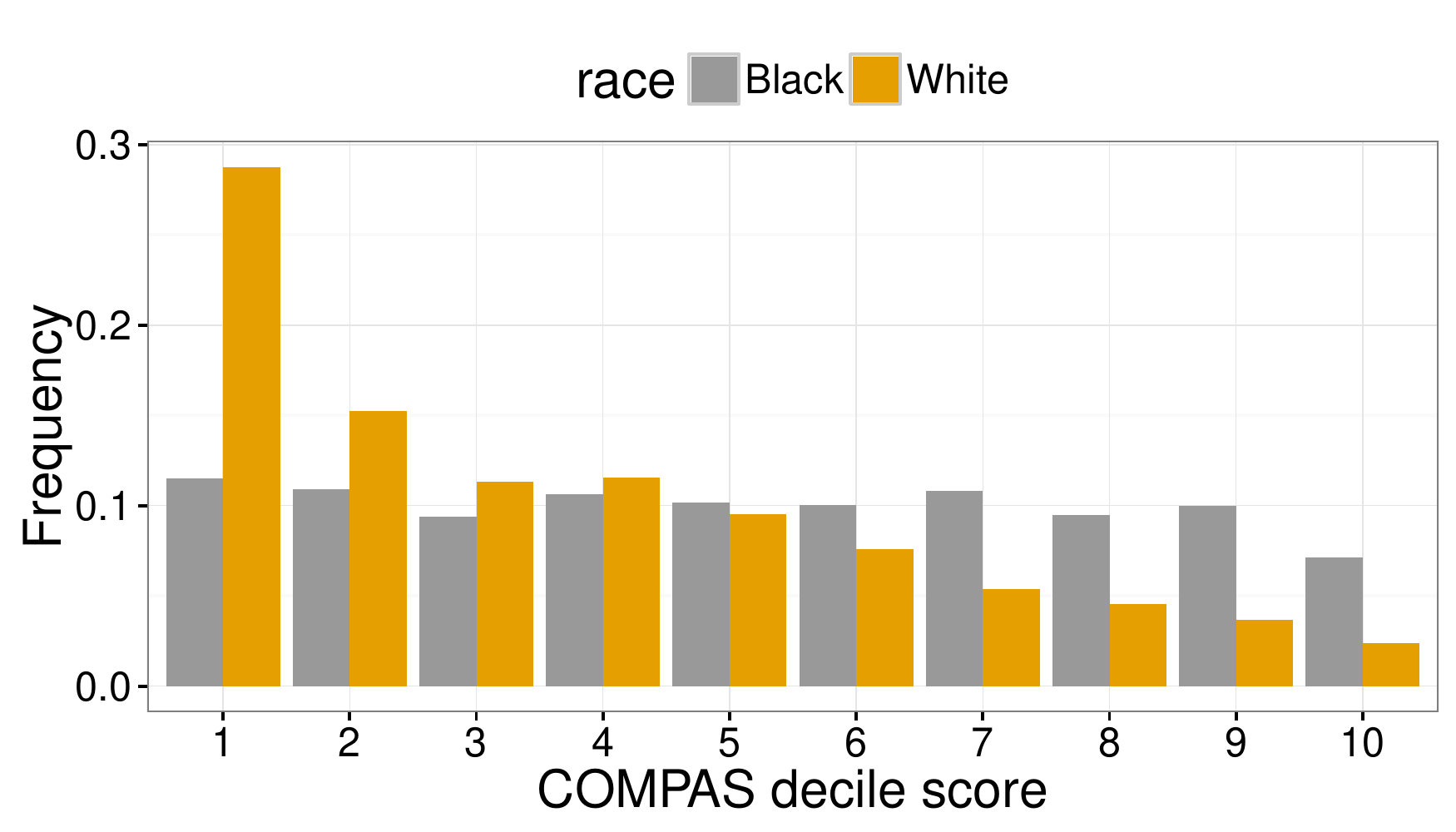}
    \caption{\small COMPAS decile score histograms for Black and White defendants.  Cohen's $d = 0.60$, non-overlap $d_{\mathrm{TV}}(f_b, f_w) = 24.5\%$.}
    \label{fig:score_plot}
 \end{figure}

  The \% non-overlap of two distributions is generally calculated assuming both distributions are normal, and thus has a one-to-one correspondence to Cohen's $d$ \cite{cohen1988}.\footnote{$d = \frac{\bar S_b - \bar S_w}{SD}$, where $SD$ is a pooled estimate of standard deviation.} Figure~\ref{fig:score_plot} shows that the COMPAS decile score is far from being normally distributed.  A more reasonable way to calculate \% non-overlap is to note that in the Gaussian case \% non-overlap is equivalent to the total variation distance.  Letting $f_{r,y}(s)$ denote the score distribution for race $r$ and recidivism outcome $y$, one can establish the following sharp bound on $\Delta$.
  
  \vspace{-0.15em}
\begin{prop}[Percent overlap bound]
  Under the MinMax policy,
  \[
  \Delta \le  (t_H - t_L) d_\mathrm{TV}(f_{b,y}, f_{w,y}).
  \]
\end{prop}
 
\vspace{-1.75em}

\section{Discussion}

\vspace{-0.5em}

The primary contribution of this paper was to show how disparate impact can result from the use of a recidivism prediction instrument that is known to be free from predictive bias.  Our analysis focussed on the simple setting where a binary risk assessment was used to inform a binary penalty policy.  While all of the formulas have natural analogs in the non-binary score and penalty setting, we find that all of the salient features are already present in the analysis of the simpler binary-binary problem.  

Our analysis indicates that there are risk assessment use cases in which it is desirable to balance error rates across different groups, even though this will generally result in risk assessments that are not free from predictive bias.  However, balancing error rates overall may not be sufficient, as this does not guarantee balance at finer levels of granularity.  That is, even if $\fpr_b = \fpr_w$, we may still see differences in error rates within prior record score categories (see e.g., Figure \ref{fig:fpr_prior}).  One needs to decide the level of granularity at which error rate balance is desirable to achieve.  

In closing, we would like to note that there is a large body of literature showing that data-driven risk assessment instruments tend to be more accurate than professional human judgements \cite{meehl1954clinical, grove2000clinical}, and investigating whether human-driven decisions are themselves prone to exhibiting racial bias \cite{anwar2012testing, sweeney1992influence}.  We should not abandon the data-driven approach on the basis of negative headlines.  Rather, we need to work to ensure that the instruments we use are demonstrably free from the kinds of quantifiable biases that could lead to disparate impact in the specific contexts in which they are to be applied.

\newpage 
\bibliographystyle{unsrtnat}
\bibliography{recidivism}
\end{document}